\documentclass{article}
\usepackage{ijcai19}

\usepackage{times}
\usepackage{soul}
\usepackage{url}
\usepackage[hidelinks]{hyperref}
\usepackage[utf8]{inputenc}
\usepackage[small]{caption}
\usepackage{graphicx}
\usepackage{amsmath}
\usepackage{booktabs}
\usepackage{tikz}
\usepackage{pgfplots}
\usepackage[detect-all,per-mode=symbol,binary-units=true]{siunitx}
\usepackage{algorithm}
\usepackage[noend]{algpseudocode}
\usepackage{multirow}
\usepackage{tablefootnote}
\usepackage{cleveref}
\usepackage{vector}
\usepackage[keeplastbox]{flushend}

\renewcommand{\vec}[1]{\vect{#1}}
\newcommand{\mat}[1]{\vect{#1}}

\pgfplotsset{every axis legend/.append style={legend cell align=left}}

\newcommand{\citet}[1]{\citeauthor{#1}~\shortcite{#1}}

\title{Satellite Image Tasking Under Orbit Prediction Uncertainty}

\author{
Duncan Eddy \And
Mykel Kochenderfer
\affiliations
Department of Aeronautics and Astronautics\\
Stanford University\\
\emails
deddy@stanford.edu,
mykel@stanford.edu
}

\begin{document}

\maketitle

\begin{abstract}

Small satellites have proven to be viable Earth observation platforms. These satellites operate in regimes of increased trajectory uncertainty where traditional planning approaches can lead to sub-optimal task plans, limiting science return. Previous formulations of the space mission planning problem decouple trajectory prediction and planning, which leads to task plans that are less robust to uncertainty. We present a Markov decision process formulation of the problem that accounts for uncertainties by incorporating a distribution of possible collection windows characterized through Monte Carlo simulation. An approximate solution technique yields tasking schedules with rewards comparable to the conventional methods while simultaneously reducing the variations caused by uncertainties and improving runtime.


\end{abstract}

\section{Introduction}

In recent years, small satellites have demonstrated their value as Earth observation platforms capable of hosting a variety of metrology payloads. Dividing the workload of a single large satellite between multiple smaller satellites 
has been shown to reduce cost, provide robustness to single-point-failures, and increase diversity in viewing geometries \cite{brown2006value,derrico2012distributed}. These small satellite platforms typically operate in low Earth orbit (LEO). Operating in this orbit regime presents its own unique set of challenges. A satellite in LEO can only be directly commanded for brief periods over ground stations so task schedules must be generated that are autonomously executed. Furthermore, drag, the most challenging astrodynamic force to model, becomes the dominant source of uncertainty in trajectory prediction. This paper studies the effect of orbit prediction errors on task scheduling. 


Satellite task scheduling involves choosing imaging opportunities to maximize an observation objective given a set of target locations to observe, the satellite trajectory, and a planning horizon. Common objectives include maximizing the number of images collected, timeliness of data return, or total monetary reward. 
There are often constraints on ground or on-board resources. These constraints are particularly relevant to small satellites as the reduction in platform size is associated with reductions in power generation, energy storage, and on-board data storage capacity.
Time constrained task-sequencing has been shown to be NP-complete \cite{garey1979computers}; consequently, various formulations and heuristics have been introduced over the years \cite{hall1994maximizing,lemaitre2002selecting,bianchessi2005earth,augenstein2016optimal}.


Past work on satellite task planning has decoupled the problems of orbit prediction and task planning. The trajectory is assumed to always be accurate when computing imaging opportunities. This modeling assumption neglects the uncertainties introduced in both the orbit determination and orbit prediction stages. Orbit prediction in low Earth orbit is especially challenging given the large uncertainties in atmospheric drag models \cite{vallado2014critical}. These modeling errors have been observed to introduce orbit prediction errors as large as \SI{200}{\kilo\meter} after a 2-day period \cite{riesing2015orbit}. 

The challenge of task planning is compounded by the fact that constraints on spacecraft agility (slew time required to reorient the sensor) encourage optimization algorithms to favor imaging at the edge of target visibility to maximize the total number of collections taken. Planning collections at the edge of feasibility leads to the possibility that a location may not be visible at the planned collection time if the true trajectory diverges from the predicted. Therefore, a schedule will tend to underperform the computed reward found by scheduling algorithms that do not account for the uncertainties introduced by orbit prediction errors.

This paper studies how uncertainty in trajectory prediction affects satellite task planning and presents a solution to the problem. The effect of orbit estimation and prediction errors on target viewing times is quantified using a high-fidelity orbit propagator. We present a formulation of the satellite tasking problem as a Markov decision process that incorporates uncertainty in transitions and solved using forward search. The performance of the MDP approach is compared to two other planning approaches in simulation.

\section{Orbit Modeling and Prediction Errors}
\label{sec:orbit_modeling}

A scheduling algorithm takes as input a set of geographic locations to be imaged and a satellite trajectory. The trajectory is used to determine the time windows when the satellite has direct line-of-sight to an imaging location. The system must decide at each point in time whether to capture an image or not. Limitations on small satellite power generation and data capacity prevent continuous collection of data for most sensor payloads. Errors in the satellite state estimate can cause the start and end times of a visibility window to shift earlier or later in time compared to the ``true'' trajectory, effectively changing the set of feasible actions. Planning algorithms either assume it is always best to capture images at the first possible opportunity to fit the greatest number of collects within the planning horizon \cite{bianchessi2008planning,aldinger2013planning}, or the behavior naturally arises when constraints on resources are introduced. The tendency to plan to the edge of feasibility gives rise to task schedules that are not robust to errors in the trajectory estimate. To understand the effect of trajectory uncertainties on plan optimality, we introduce a high-fidelity orbit dynamics model for trajectory and image opportunity prediction then analyze how small differences in initial conditions affect the start and end times. 

\subsection{Orbit Modeling}

The equations of motion describing a satellite trajectory can be formulated in the Cartesian Earth-centered inertial frame from Newtonian mechanics. A satellite's acceleration is related to the total force $\vec{f}_{\text{total}}$ acting on it, which is a function of time $t$, 
position $\vec{r}$, and velocity $\vec{v}$. 
By numerically integrating
\begin{equation}
	\ddot{\vec{r}} = \frac{1}{m}\vec{f}_{\text{total}}(t, \vec{r}, \vec{v})
\label{eqn:newton}
\end{equation}
forward in time, the position and velocity at future times can be computed.

To simplify 
the formulation of the force model, the total force can be decomposed into the sum of Earth's gravity and other perturbing forces as
\begin{equation}
	\vec{f}_{\text{total}} = \vec{f}_{\text{grav}} + \vec{f}_{\text{drag}} + \vec{f}_{\text{srp}} + \vec{f}_{\text{tb}} + \vec{f}_{\text{rel}} + \vec{f}_{\text{other}}
\label{eqn:forces}
\end{equation}
The most significant perturbing forces outside of gravity are atmospheric drag, solar radiation pressure, third-body gravity of the Sun and Moon, and first-order corrections for the effects of special and general relativity.

The fidelity of the dynamics simulation is determined by which additional perturbing forces are included and how closely the individual perturbation models capture the effect of the true physical phenomenon they are describing. Some forces like gravity and relativity are well understood, while others like atmospheric drag are notoriously difficult to model. The instantaneous drag force depends on the local atmospheric density, which varies with geocentric position, solar activity, altitude, and the molecular interactions of the atomic gases with the spacecraft surface materials. The selection of models used in the rest of this work are shown in \Cref{tab:orb_models} \cite{montenbruck2012satellite,vallado2001fundamentals,petit2010iers,pavlis2008earth}. The dynamics modeling code is available as the open-source Julia package available at \url{https://github.com/sisl/SatelliteDynamics.jl}.
\begin{table}
\centering
\caption{Orbit models used in dynamics simulations}
\begin{tabular}{ll}
\toprule
Orbit Perturbation & Model Description \\
\midrule
Gravity Field & EGM2008 (up to 120 $\times$ 120) \\
Atmospheric Drag & Harris-Priester \\
\phantom{-} & Cannon-ball spacecraft model \\
Solar Radiation Pressure & Flat-plate spacecraft model \\
\phantom{-} & Conical eclipse model \\
Third-body gravity & Analytical Sun position \\
\phantom{-} & Analytical Moon position \\
Relativistic corrections & Corrections for special \\
\phantom{-} & and general relativistic effects \\
Reference Frames & IAU 2010 precession/nutation \\
\phantom{-} & IERS C04 14 \\
\phantom{-} & Earth orientation parameters \\
Numerical Integration & RK4 \\
\bottomrule
\end{tabular}
\label{tab:orb_models}
\end{table}

\subsection{Effect Prediction Errors on Image Collection} 

Let there be a set of \emph{images} $I$, which have been requested to be imaged over a time span $[0, T]$ where $T$ is the planning horizon. Each image, $i \in I$, is defined by an Earth-fixed center point. Over the planning horizon each image has a set of windows of \emph{opportunity} $O_i$. Each individual opportunity for image $i$ is denoted $o_i \in O_i$ and has a start time $t_s$ and end time $t_e$ during which there is a direct line of sight from the satellite to image center. The visibility periods can be further constrained by other geometric considerations, such as a maximum allowable look-angle, $\theta_{max}$, from the off-nadir direction to the image center.

The imaging opportunities are found by taking an initial state estimate then predicting the trajectory forward in time using the orbit dynamics model. A search algorithm is used to find the periods of the trajectory where there is a direct line of sight vector from the satellite to the image center and all other geometric imaging constraints are met. The probability that an opportunity exists, $p(o_i \mid \mat{\Sigma}_{\text{orbit}},\vec{U}_{\text{force}})$, is then conditioned on the initial orbit estimate, with covariance $\mat{\Sigma}_{\text{orbit}}$, and on any additional stochastic force model parameters $\vec{U}_{\text{force}}$. 

If the dependency on force parameters is neglected, then the satellite trajectory, and consequently the imaging opportunities, is entirely predetermined by the error in the initial conditions. \Cref{tab:orb_errors} presents various types of orbit determination methods currently employed by small satellite missions and the typical standard deviation of the position associated with each. The effect of orbit estimation errors on imaging opportunities can be characterized through Monte Carlo simulation of different trajectories.

\begin{table}
\caption{Initial position error for different orbit determination techniques}
\centering
\begin{tabular}{ccc}
\toprule
 Case & Position S.Dev [m] & Sensor System\\
\midrule
1 & 1 & GPS \\
2 & 10 & GPS \\
3 & 100 & GPS \\
4 & 1000 & COM\footnotemark \\
5 & 2500 & TLE \\
6 & 5000 & TLE \\
\bottomrule
\end{tabular}
\label{tab:orb_errors}
\end{table}
\footnotetext{COM is shorthand for performing orbit determination based on time-of-flight analysis to communications signal with a satellite.}

\begin{figure}
\begin{tikzpicture}[]
\begin{axis}[legend pos = {north west}, ylabel = {Standard deviation of oppotrunity window [s]}, xmin = {0}, xmax = {24}, xlabel = {Time since initial epoch [hr]}, ymin = {0}]\addplot+ coordinates {
(1.0, 0.0)
(2.0, 0.0)
(3.0, 0.0)
(4.0, 0.0)
(5.0, 0.0828539098642126)
(6.0, 0.0)
(7.0, 0.2107881486829658)
(8.0, 0.13140153920440992)
(9.0, 0.098530571668512)
(10.0, 0.08612819873644528)
(11.0, 0.10444130910722)
(12.0, 0.05679618342470662)
(13.0, 0.0)
(14.0, 0.12812591191634937)
(15.0, 0.0)
(16.0, 0.09910807642422531)
(17.0, 0.1590714573062532)
(18.0, 0.12872609869300533)
(19.0, 0.11825208725998053)
(20.0, 0.14035131799278383)
(21.0, 0.14307210005542867)
(22.0, 0.11970180261300997)
(23.0, 0.14430964630530957)
(24.0, 0.06811990105241413)
};
\addlegendentry{1 m}
\addplot+ coordinates {
(1.0, 0.0)
(2.0, 0.10573528406248887)
(3.0, 0.19335284094348648)
(4.0, 0.22249362693918612)
(5.0, 0.2636596174555254)
(6.0, 0.35561681782677196)
(7.0, 0.3401002833944373)
(8.0, 0.29502185258643165)
(9.0, 0.3375156088637766)
(10.0, 0.3654903624597742)
(11.0, 0.41162446536339364)
(12.0, 0.39860912793402475)
(13.0, 0.43416685779890246)
(14.0, 0.42876211408310094)
(15.0, 0.4367538989568008)
(16.0, 0.5250820341078382)
(17.0, 0.45122372362735835)
(18.0, 0.5204219646756001)
(19.0, 0.4965202455422646)
(20.0, 0.49682408441446474)
(21.0, 0.5480239529363384)
(22.0, 0.5621399780685763)
(23.0, 0.5692368394172415)
(24.0, 0.5679358170766582)
};
\addlegendentry{10 m}
\addplot+ coordinates {
(1.0, 0.13987236879147186)
(2.0, 0.302451277937322)
(3.0, 0.48193204887221036)
(4.0, 0.5588851188211474)
(5.0, 0.661519405723312)
(6.0, 0.7841817733639523)
(7.0, 0.8322735168885794)
(8.0, 0.9717052757763878)
(9.0, 0.9892535916435301)
(10.0, 1.15259359265383)
(11.0, 1.3054020353233056)
(12.0, 1.3744913360300965)
(13.0, 1.503989580959811)
(14.0, 1.6300406023797245)
(15.0, 1.7517208067364398)
(16.0, 1.8780036447698938)
(17.0, 1.9172353868872525)
(18.0, 2.0681638175519312)
(19.0, 2.1936644714748277)
(20.0, 2.3130827392884106)
(21.0, 2.4403528271381343)
(22.0, 2.5820321231212118)
(23.0, 2.6409253125375205)
(24.0, 2.7865471882397874)
};
\addlegendentry{100 m}
\addplot+ coordinates {
(1.0, 0.4761344927854081)
(2.0, 1.0720875414788702)
(3.0, 2.198184626071436)
(4.0, 2.863087280743819)
(5.0, 3.9117509693174295)
(6.0, 5.099789228372612)
(7.0, 5.619624715335303)
(8.0, 6.89976798135901)
(9.0, 7.410644173000314)
(10.0, 8.570500106433334)
(11.0, 9.626637185063416)
(12.0, 10.426092926763452)
(13.0, 11.545401907843287)
(14.0, 12.612943261939817)
(15.0, 13.314406841393113)
(16.0, 14.496677053269696)
(17.0, 14.940646244949326)
(18.0, 16.033287762102315)
(19.0, 17.290349744753577)
(20.0, 18.1282022729607)
(21.0, 19.051407507524523)
(22.0, 20.211110520775136)
(23.0, 21.076477664240883)
(24.0, 21.900659326844636)
};
\addlegendentry{1000 m}
\addplot+ coordinates {
(1.0, 0.5958585861614573)
(2.0, 1.5930582901924393)
(3.0, 3.252505365020528)
(4.0, 4.572727699039086)
(5.0, 5.812789944139508)
(6.0, 7.733771412901478)
(7.0, 8.796188113919172)
(8.0, 10.463174879125672)
(9.0, 11.6273807361485)
(10.0, 13.100495803848359)
(11.0, 14.868484113018635)
(12.0, 16.456480583956804)
(13.0, 17.731087811988328)
(14.0, 19.317006522444625)
(15.0, 21.141303378338343)
(16.0, 22.16891675881983)
(17.0, 23.454645641715985)
(18.0, 24.73710196594863)
(19.0, 26.703889435858425)
(20.0, 27.988032250206132)
(21.0, 29.443924522743693)
(22.0, 31.259829585058533)
(23.0, 33.08677608491366)
(24.0, 33.76395411755295)
};
\addlegendentry{2500 m}
\addplot+ coordinates {
(1.0, 1.3046095353359144)
(2.0, 4.357032305678085)
(3.0, 8.046233301125456)
(4.0, 10.136635570513151)
(5.0, 13.829004720798691)
(6.0, 18.18272066861099)
(7.0, 19.687497782015523)
(8.0, 24.2693112339468)
(9.0, 26.214668554864758)
(10.0, 30.136791060957144)
(11.0, 33.8369734576197)
(12.0, 36.92613302232753)
(13.0, 40.40766424974587)
(14.0, 44.064271275185355)
(15.0, 47.05668314460155)
(16.0, 50.46180998093664)
(17.0, 53.283496657434)
(18.0, 56.14253323888328)
(19.0, 60.17873713010119)
(20.0, 63.51149854759525)
(21.0, 67.24525932523612)
(22.0, 70.19443677029953)
(23.0, 74.06444818145667)
(24.0, 76.04384180740479)
};
\addlegendentry{5000 m}
\end{axis}

\end{tikzpicture}	
\caption{Effect of orbit determination and trajectory prediction errors on start time for opportunity window start time. \SI{550}{\kilo\meter} altitude orbit, 600 random locations.}
\label{fig:window_errors}
\end{figure}
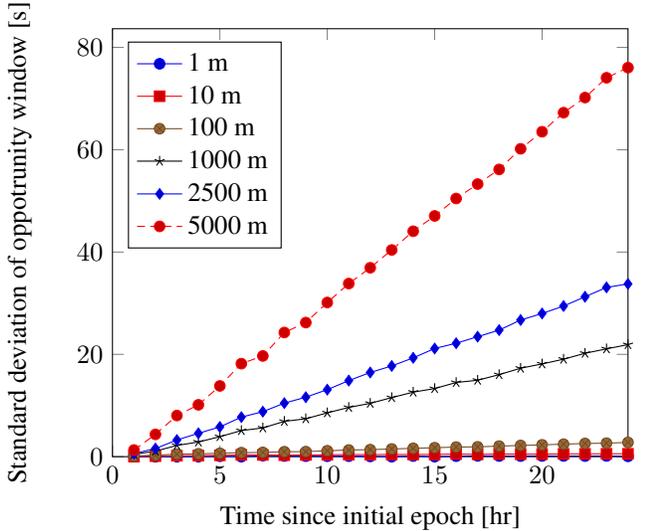

\Cref{fig:window_errors} shows how the start of opportunity windows for 600 targets varies depending on the initial orbit uncertainty. As expected, when the initial uncertainty is small (Cases 1--3), there is very little variation in the opportunity window start times because there are no significant deviation in the trajectories. However, when the initial uncertainty is large (Cases 4--6), the predicted trajectories diverge significantly. As a result, there is significant variation in the opportunity start times.

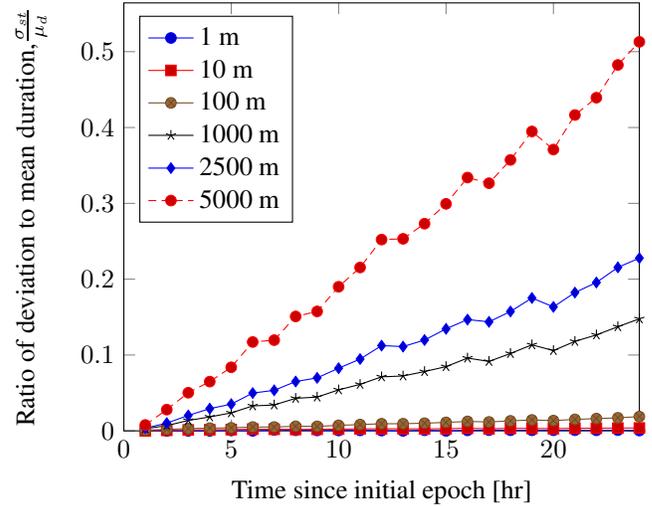
\begin{figure}
\begin{tikzpicture}[]
\begin{axis}[legend pos = {north west}, ylabel = {Ratio of deviation to mean duration,$\frac{\sigma_{st}}{\mu_d}$}, xmin = {0}, xmax = {24}, xlabel = {Time since initial epoch [hr]}, ymin = {0}]\addplot+ coordinates {
(1.0, 0.0)
(2.0, 0.0)
(3.0, 0.0)
(4.0, 0.0)
(5.0, 0.0005022312744938161)
(6.0, 0.0)
(7.0, 0.001280578035528636)
(8.0, 0.0008167888795482008)
(9.0, 0.0005919631587237735)
(10.0, 0.000543051592324896)
(11.0, 0.0006649765392040191)
(12.0, 0.0003879717133379694)
(13.0, 0.0)
(14.0, 0.0007943052138352359)
(15.0, 0.0)
(16.0, 0.0006560504424335455)
(17.0, 0.0009748468963064359)
(18.0, 0.000819149928974637)
(19.0, 0.0007757901035467063)
(20.0, 0.0008196902996451748)
(21.0, 0.0008861210393672222)
(22.0, 0.0007491728798376543)
(23.0, 0.0009401841241959953)
(24.0, 0.0004595337446702142)
};
\addlegendentry{1 m}
\addplot+ coordinates {
(1.0, 0.0)
(2.0, 0.000679141078270122)
(3.0, 0.0012079565582351337)
(4.0, 0.0014250920111653364)
(5.0, 0.0015982119120782302)
(6.0, 0.002291550321776757)
(7.0, 0.0020661738124899453)
(8.0, 0.001833848902191694)
(9.0, 0.0020277646070476327)
(10.0, 0.002304473171679098)
(11.0, 0.002620807942459349)
(12.0, 0.0027228777884649963)
(13.0, 0.0027208607500432054)
(14.0, 0.0026580726538248936)
(15.0, 0.002778330108555299)
(16.0, 0.0034758045279360376)
(17.0, 0.002765260682003393)
(18.0, 0.0033117108319854894)
(19.0, 0.0032574096713862345)
(20.0, 0.0029015893006830943)
(21.0, 0.003394201626913604)
(22.0, 0.003518242975864249)
(23.0, 0.0037086047470127055)
(24.0, 0.0038312691111039915)
};
\addlegendentry{10 m}
\addplot+ coordinates {
(1.0, 0.0008367842182998418)
(2.0, 0.0019426541371103234)
(3.0, 0.0030108323012902317)
(4.0, 0.003579710254842055)
(5.0, 0.004009898081856421)
(6.0, 0.005053169324401816)
(7.0, 0.005056219678093385)
(8.0, 0.006040097157597336)
(9.0, 0.005943350078778506)
(10.0, 0.007267280576823173)
(11.0, 0.00831147881177952)
(12.0, 0.009389077336767295)
(13.0, 0.00942528464759712)
(14.0, 0.010105291973091735)
(15.0, 0.011143251773511949)
(16.0, 0.012431530976035995)
(17.0, 0.011749505524410852)
(18.0, 0.013160782945002232)
(19.0, 0.014391485401273987)
(20.0, 0.013509039393337812)
(21.0, 0.015114393251854218)
(22.0, 0.01616006107916214)
(23.0, 0.01720575246080254)
(24.0, 0.018797920201422624)
};
\addlegendentry{100 m}
\addplot+ coordinates {
(1.0, 0.0028484670188506987)
(2.0, 0.006886052232948298)
(3.0, 0.013732984332258844)
(4.0, 0.01833833565117049)
(5.0, 0.023711659208870868)
(6.0, 0.03286240431115828)
(7.0, 0.034140287408643125)
(8.0, 0.04288879561654055)
(9.0, 0.04452250969968864)
(10.0, 0.05403832656551159)
(11.0, 0.0612926813558454)
(12.0, 0.07122008720145258)
(13.0, 0.07235335984368209)
(14.0, 0.07819282177135008)
(15.0, 0.08469716582577319)
(16.0, 0.0959614164430409)
(17.0, 0.09156163442106532)
(18.0, 0.10202800104179403)
(19.0, 0.11343294253530416)
(20.0, 0.10587368729886452)
(21.0, 0.11799542339446896)
(22.0, 0.12649446827896418)
(23.0, 0.1373142419874512)
(24.0, 0.14774084864668052)
};
\addlegendentry{1000 m}
\addplot+ coordinates {
(1.0, 0.0035647145004150693)
(2.0, 0.010232263851573402)
(3.0, 0.020319769635657157)
(4.0, 0.029288738750778753)
(5.0, 0.03523508916832658)
(6.0, 0.04983545625903206)
(7.0, 0.053438513338834745)
(8.0, 0.06503884914729464)
(9.0, 0.06985629852439044)
(10.0, 0.08260064892678483)
(11.0, 0.09466745671039824)
(12.0, 0.11241334509975885)
(13.0, 0.11111815657185442)
(14.0, 0.11975406665971305)
(15.0, 0.13448653772853666)
(16.0, 0.14674815789625364)
(17.0, 0.14373847385941613)
(18.0, 0.1574148173850175)
(19.0, 0.17519025355551104)
(20.0, 0.1634578062375575)
(21.0, 0.18236176718616426)
(22.0, 0.19564464395899814)
(23.0, 0.21556190034622108)
(24.0, 0.2277700940665494)
};
\addlegendentry{2500 m}
\addplot+ coordinates {
(1.0, 0.007804805764318645)
(2.0, 0.02798535649071692)
(3.0, 0.05026820520328045)
(4.0, 0.06492607707627096)
(5.0, 0.0838265650624131)
(6.0, 0.11716717913838666)
(7.0, 0.11960528801876229)
(8.0, 0.1508574682625689)
(9.0, 0.15749546298018424)
(10.0, 0.1900171211439749)
(11.0, 0.21543959664357293)
(12.0, 0.25224045405463236)
(13.0, 0.25322897333858974)
(14.0, 0.27317253703201744)
(15.0, 0.29934248990488005)
(16.0, 0.3340342579376195)
(17.0, 0.32654036255448304)
(18.0, 0.3572636208354723)
(19.0, 0.3948012233122836)
(20.0, 0.3709246198747424)
(21.0, 0.41648538787584294)
(22.0, 0.4393231112300046)
(23.0, 0.48253335885960436)
(24.0, 0.5129882874899991)
};
\addlegendentry{5000 m}
\end{axis}

\end{tikzpicture}	
\caption{Ratio of standard deviation of the start time to total opportunity duration for a \SI{550}{\kilo\meter} altitude orbit and 600 random locations.}
\label{fig:coef_variation}
\end{figure}

\Cref{fig:coef_variation} shows the ratio of the standard deviation of the opportunity start times to the mean opportunity duration for a \SI{550}{\kilo\meter} orbit. The mean duration of the opportunities is approximately \SI{180}{\second} at this altitude. Given that collection times are typically on the order of a few seconds ($\leq$\SI{10}{\second}), a ratio of 0.1 means that it is possible a collect taken at the edge of an opportunity window to not be feasible. For Case 6, the ratio becomes as large as 0.52 after 24 hours, which presents a significant challenge for ensuring collects are feasible when these uncertainties are not accounted for.

While having a more accurate navigation solution significantly reduces the variation in the imaging windows, and consequently ameliorates the effect of uncertainty on task plans, more accurate navigation methods are not available to all small satellite missions. Many missions, especially cubesats, rely solely on NORAD Two Line Elements (TLEs), and therefore being able to generate task plans that are more robust to the highest levels of uncertainty would still greatly enhance mission planning capabilities.

\section{Satellite Tasking as an MDP}

This section poses the satellite image tasking problem as an MDP and introduces an approximate online solution approach. In the MDP model the agent (satellite) chooses an action in the current state, receives a reward, then transitions probabilistically to the next state. The process is repeated over the planning horizon. The rest of the section describes the state, action, transition, and rewards of the MDP as well as the solution approach.

\subsubsection{Action}

The action the satellite can take is to choose the next image collection. For each image $i$ the set of opportunities for each image $O_i$ is discretized into a set of possible \emph{collects} $C_i$ to further simplify the planning problem. Although an image could be captured any time over the continuous period $[t_s,t_e]$, this opportunity window is discretized into collect sub-intervals to simplify formulating the problem and implementing the solution. 

The set of possible actions can be limited by constraints on spacecraft on-board resources and capabilities. \Cref{alg:action_generation} is used to generate the set of possible actions for a given state $s_t$. In \Cref{alg:action_generation}, $s^p_{t+1}$ refers to \emph{possible} future states, $h$ is the planning horizon beyond which possible actions are neglected, and $c^p_{t+1}$ is the collect that starts at $s^p_{t+1}$. The advantage of this approach is that the planing process can impose arbitrary constraints on image collections or spacecraft resources provided they can be formulated as a function of the current state and future state $f_c(s_t,s^p_{t+1})$. The inclusion of the planning horizon, $h$ also serves to limit the size of the decision space to reduce the computational complexity.

The constraint used in this work is an agility constraint imposing that the spacecraft be pointed at the image center at the start and end of each collect, and that slews between pointing vectors are limited to a \SI{1}{\degree\per\second} maximum slew rate.

\begin{algorithm}[tb]
\caption{Action Space Generation}
\label{alg:action_generation}
\begin{algorithmic}[1] 
\State Let $A(s)=\{\textsc{nil}\}$.
\For{$s^p_{t+1} \in S^p$}
\If{$s^p_{t+1} - s_t > h$}
\State{\textbf{continue}}
\EndIf
\If{$f_c(s_t, s^p_{t+1})$ $\forall$ $f_c \in F_c$}
\State {$A(s) \gets A(s) \cup \{c^p_{t+1}\}$}
\EndIf
\EndFor
\State \textbf{return} $A(s)$
\end{algorithmic}
\end{algorithm}

\subsubsection{State}

The state $s$ for time $t$ consists of the time and a list of Boolean indicators $b_i$, each indicating whether image $i$ has been already collected. The state has dimension $1 + N$, where $N$ is the number of images requested for collection. The upper bound on the size of the state space is $2^NP$, where $P$ is the number of discretized collections. Because the size of the state space grows exponentially with the number of images taken, finding an exact solution quickly becomes computationally intractable even for small problem sizes.

This choice of spacecraft state has the benefit of being easily extended to account for on-board resources like energy storage, thermal capacity, or momentum storage by augmenting the state with the appropriate variables.

\subsubsection{Reward}

The reward for each state $s$ is
\begin{equation}
	R(s) = \sum_{i \in I}r_iD(b_i) \\
\label{eqn:mdp_reward}
\end{equation}
where $r_i$ is the reward for collecting image $i \in I$, and $D(b_i)$ is the indicator variable 
\begin{equation}
  D(b_i) =
  \begin{cases}
   +1 & \text{if $b_i=1$} \\
   -1 & \text{if $b_i=0$} \\
  \end{cases}
\label{eqn:mdp_indicator}
\end{equation} 
\Cref{eqn:mdp_reward} is formulated such that the agent is rewarded for every unique image collected and penalized for each image missing that has not yet been collected.

\subsubsection{Transition}

The state of the agent evolves probabilistically given the current state and the action taken. The distribution over the future state $s_{t+1}$ is conditioned on the current state $s_t$ and the action $a$ taken at $t$. However because time advances deterministically, regardless of whether the image collect is successful or not, the probability that the state transitions from $b_i = 0$ to $b_i = 1$ for action $a$ is equal to the probability the associated collect is feasible given the initial orbit uncertainty. That is
\begin{equation}
  p(s_{t+1} \mid s_t,a) = p(c_i \mid \mat{\Sigma}_{\text{orbit}})
\label{eqn:mdp_transition}
\end{equation}
The probability a collect is feasible is equal to the probability the collect start and end times are entirely within a single opportunity. The distribution $p(c_i \mid \mat{\Sigma}_{\text{orbit}})$ is equivalent to
\begin{equation}
\resizebox{.91\linewidth}{!}{$
  p(c_i \mid \mat{\Sigma}_{\text{orbit}}) = p(o_{{s}} < c_{{s}} < o_{{e}}, o_{{s}} < c_{{e}} < o_{{e}}  \mid  \mat{\Sigma}_{\text{orbit}})
 $}
\label{eqn:mdp_collect}
\end{equation}
Subscripts $s$ and $e$ denote the start and end time of opportunity $o_i$ or collect $c_i$ for image $i$.

\Cref{sec:orbit_modeling} shows how the probability distribution ${p(c_i~\mid~\mat{\Sigma}_{\text{orbit}})}$ can be characterized using Monte Carlo simulation. The characterization of the distribution can occur offline before solving the MDP to improve runtime performance.


\subsubsection{Forward Search}

An MDP may be solved using a dynamic programming algorithm such as value iteration or policy iteration \cite{kochenderfer2015decision}. Solving our problem with these approaches is intractable due to the exponential growth in the state space with the number of images being planned, $N$. Instead we consider the approximate solution provided by the forward search algorithm presented in \Cref{alg:forward_search}.

To find the tasking plan given the initial satellite state, the  algorithm performs a search for the optimal action up to a maximum look-ahead horizon of $d$. After the optimal action is found, it is added to the tasking plan. The state is transitioned forward in time assuming the collection was successful, i.e. $p(c_i \mid \mat{\Sigma}_{\text{orbit}}) = 1$. The process is repeated until the end of the planning period $T$. 

The horizon $d$ is chosen based on the number of images in the problem, computational resources, and the desired optimality of the solution. Larger values of $d$ will lead to better plans due to greater ``foresight'' of the algorithm when choosing actions at the cost of increased runtime. The computational complexity is $O((|S|\times|A|)^d)$. Values for $d$ of 1, 2, or 3 were found to work well in practice depending on the number of images being requested. Values greater than 4 were not easily solvable.

\begin{algorithm}[tb]
\caption{Satellite Tasking MDP Forward Search}
\label{alg:forward_search}
\hspace*{\algorithmicindent} \textbf{Input} \\
\hspace*{\algorithmicindent}\hspace*{\algorithmicindent} $s$, state \\
\hspace*{\algorithmicindent}\hspace*{\algorithmicindent} $d$, search depth \\
\hspace*{\algorithmicindent}\hspace*{\algorithmicindent}$A(s)$, action space \\
\hspace*{\algorithmicindent}\hspace*{\algorithmicindent} $p(c_i \mid \mat{\Sigma}_{\text{orbit}})$, collect probability distribution \\
\hspace*{\algorithmicindent} \textbf{Output} \\
\hspace*{\algorithmicindent}\hspace*{\algorithmicindent} $\pi(s)$, Policy
\begin{algorithmic}[1]
\Function{SelectAction}{$s,d,p(c \mid \Sigma),\gamma$}
\If{$d = 0$}
	\State \Return (\textsc{nil}, 0)
\EndIf
	\State $(a^\star, v^\star) \gets (\texttt{NIL}, -\infty)$
\For{$a \in A(s)$}
	\State $v \gets R(s)$
	\For{$s^\prime \in S(s, a)$}
		\State $(a^\prime, v^\prime) \gets \textsc{SelectAction}(s^\prime, d-1,p(c \mid \Sigma)$
		\State $v \gets v + p(c \mid \Sigma)v^\prime$
	\EndFor
	\If{$v > v^\star$}
		\State $(a^\star, v^\star) \gets (a, v)$
	\EndIf
\EndFor
\State \Return $(a^\star, v^\star)$
\EndFunction
\end{algorithmic}
\end{algorithm}

\section{Experimental Results}

The simulations require a set of hypothetical imaging targets that must be collected. We used Landsat image locations to provide a realistic set of  candidate imaging points \cite{nag2018scheduling}. The Landsat program regularly collects images of the entire globe along the pre-defined WRS-2 grid. The data set contains 16,896 coastal or land images over a 16-day repeat cycle. \Cref{tab:test_cases} presents the three different test cases used to compare the performance of each of the planning algorithms. The planning horizon is assumed to be 1 day long (a natural planning duration) and locations were randomly selected out of the total 16,896.

\begin{table}[t]
\caption{Test cases.}
\centering
\begin{tabular}{ccc}
\toprule
 Case & Number of Images & Position Uncertainty [m]\\
\midrule
1 & 600 & 1000 \\
2 & 600 & 2500 \\
3 & 600 & 5000 \\
4 & 1200 & 1000 \\
5 & 1200 & 2500 \\
6 & 1200 & 5000 \\
\bottomrule
\end{tabular}
\label{tab:test_cases}
\end{table}

\subsection{Alternative Solutions}

To evaluate the performance of the MDP model for satellite task scheduling we implemented two other formulations of the tasking problem. The first is graph-based method where the tasking schedule is generated by finding the longest weighted path through a decision tree based on the work of \citet{augenstein2014optimal}. The second is a mixed-integer linear programming (MILP) formulation based on the work of \citet{nag2018scheduling}.

The first approach constructs a graph where each possible collect is a node and feasible transitions between collects are edges. The optimal plan is then simply the longest weighted path, where the weights are the rewards for collecting each image. \Cref{alg:graph_solution} allows for the direct calculation of this cost. Here, $R_i$ is the total reward collected for the plan found passing through node $i$, $r_i$ is the reward for collecting the image associated with node $i$, and $n_j$ is the prior node associated with the highest reward. The optimal path can be found by iterating through the graph from the start to end with  \cref{alg:graph_solution} then back-tracking from  node $\text{argmax}_i R_i$.

\begin{algorithm}[h]
\caption{Graph-based Tasking Algorithm}
\label{alg:graph_solution}
\begin{algorithmic}[1] 
\State $R_j = 0$
\ForAll{$i  \mid  \text{Edge}(i,j)$}
\If{$R_i + r_j > R_j$}
\State $R_j \gets R_i + R_j$
\State $n_j \gets i$
\EndIf
\EndFor
\end{algorithmic}
\end{algorithm}

The MILP formulation attempts to solve the problem
\begin{equation}
\begin{aligned}
{\text{maximize}} \hspace{0.5cm} & \sum_{i \in I}\sum_{c_j \in C_i} r_ic_j \\
\text{subject to} \hspace{0.5cm} & c_j \in \{0, 1\} \\
& \sum_{c_j \in C_i} c_j \leq 1 \; \forall i \in I \\
& c_k + c_l \leq 1 \; \forall c_k,c_l \; s.t. \; c_l \notin A(c_k)
\end{aligned}
\end{equation}
The objective represents the total reward from all image collections across all images. The first constraint is the integer constraint on the binary decision of whether the collect is taken or not. The second constraint imposes a limit of at most one collect taken per each image to ensure that the solution attempts to capture all images. The final constraint captures planning constraints on agility and time. For all collects $C$, across all images, if collect $c_l$ is not in the action space of collect $c_k$, then a mutual-exclusion constraint is added. The action space for collects $A(c)$, can be computed using a modified version of \Cref{alg:action_generation} where only the constraint function is one on spacecraft agility. The MILP formulation can only incorporate agility constraints, not resource constraints, since planning is performed over collections, not the true spacecraft state.

\subsection{Test Results}

To evaluate the performance of each of the three scheduling methods, a trajectory was simulated for a spacecraft in a \SI{550}{\kilo\meter} circular polar orbit using the force model from \cref{tab:orb_models}. Each of the three scheduling algorithms was used to compute a tasking plan based on this nominal trajectory. The MDP algorithm was also provided with a distribution of collect times computed using 10 random trajectories with initial conditions sampled from the orbit determination covariance $\Sigma_{\text{orbit}}$. Tests were run on a machine with \SI{2.6}{\giga\hertz} Intel Core i7 with \SI{16}{\giga\byte} of memory. The MILP problem was solved using Gurobi. 

For the 600 image test cases, nominal predicted reward of the Graph-based, MILP, and MDP plans were 546, 552, and 489, respectively. It is important to note that in this scenario only 552 out of the 600 targets were found to have imaging opportunities, which means that the MILP formulation was able to find an optimal policy that captures entire planning reward. The result is understandable, since the MILP formulation provides the most complete exploration of the decision space. The graph-based approach also performs well due to its more thorough search exploration of the decision space. The MDP approach performs the worst of the three when there is no uncertainty in the trajectory because it uses a small look-ahead when choosing actions compared to the other techniques that consider the entire planning space.

\Cref{tab:case_results_600} shows the results of evaluating  each plan against 100 randomized trajectories drawn from the relevant distribution. It can be seen that the mean rewards of all planning techniques closely track one another for all test cases. While it might be initially surprising that the mean reward for the graph-based approach outperforms the MDP approach even in the presence of uncertainties, it is important to remember that initial condition errors are drawn from a normal distribution, which means that the majority of the simulated trajectories will not have significant deviations from the nominal trajectory. When the trajectory deviations are small, it is expected that the alternative approaches will outperform the MDP approach as seen from the nominal case.

\begin{table}[ht]
\caption{Planning results for 600 imaging targets over 24 hours. \\ ~}
\centering
\begin{tabular}{cccc}
\toprule
 Approach & Runtime [s] & \vtop{\hbox{\strut Mean}\hbox{\strut Reward}} & \vtop{\hbox{\strut S.Dev}\hbox{\strut Reward}}\\
\midrule
\multicolumn{4}{c}{Case 1}\\
\midrule
Graph & 237.95 & 386.08 & 181.44 \\
MILP & 129.60 & 383.83 & 161.98 \\
MDP & 76.76 & 381.97 & 111.03 \\
\midrule
\multicolumn{4}{c}{Case 2}\\
\midrule
Graph & 234.92 & 379.78 & 168.76 \\
MILP & 131.69 & 358.96 & 189.04 \\
MDP & 81.24 & 367.80 & 139.57 \\
\midrule
\multicolumn{4}{c}{Case 3}\\
\midrule
Graph & 213.91 & 334.81 & 179.41 \\
MILP & 122.45 & 304.40 & 196.58 \\
MDP & 81.97 & 323.93 & 157.86 \\
\bottomrule
\end{tabular}
\label{tab:case_results_600}
\end{table}

\Cref{tab:case_results_1200} shows how increasing the number of imaging targets to 1200 affects the performance of the planning algorithms. The runtime of the non-MDP methods drastically increases with the size of the problem. The MILP formulation was not able to converge on a solution within a 12-hour time limit. Because of the bounded nature of the forward search algorithm, the MDP solution time did not grow as drastically with the number of imaging targets. The improved runtime performances indicate that a sub-optimal online approach could be of interest to mission planning when considering large sets of image requests.

In all cases, the MDP formulation has the smallest variance in the realized reward. This result shows that by including distribution of target visibility in the planning process the MDP algorithm is able to generate a tasking plan more robust to the trajectory uncertainties. A smaller variance in the schedule reward indicates that the actions selected by the MDP plan are more frequently feasible.

It is expected that there is a limit to the improvement achieved by accounting for uncertainties. After a certain level of trajectory uncertainty is reached, no precomputed plan can account for the variations present and solutions must be found entirely online.

The MDP formulation's ability to successfully generate improved imaging tasking plans in the presence of large, but not small, trajectory uncertainties suggests a number of areas for further improvement. Other approximate solution techniques such as branch-and-bound, sparse sampling, or Monte Carlo tree search may yield better solutions or improve runtime performance. It would also be worth investigating other methods of state space selection to determine whether a more compact representation of the problem is possible. 

\begin{table}
\centering
\caption{Planning results for 1200 imaging targets over 24 hours. DNC means the method did not complete before the 12-hour cutoff.}
\begin{tabular}{cccc}
\toprule
 Approach & Runtime [s] & \vtop{\hbox{\strut Mean}\hbox{\strut Reward}} & \vtop{\hbox{\strut S.Dev}\hbox{\strut Reward}}\\
\midrule
\multicolumn{4}{c}{Case 4}\\
\midrule
Graph & 34376.87 & 727.64 & 281.69 \\
MILP & $DNC$ & - & - \\
MDP & 390.41 & 657.30 & 153.89 \\
\midrule
\multicolumn{4}{c}{Case 5}\\
\midrule
Graph & 34353.98 & 670.27 & 320.95 \\
MILP & $DNC$ & - & - \\
MDP & 412.86 & 603.82 & 207.45 \\
\midrule
\multicolumn{4}{c}{Case 6}\\
\midrule
Graph & 32216.93 & 554.94 & 352.00 \\
MILP & $DNC$ & - & - \\
MDP & 385.82 & 494.21 & 259.60 \\
\bottomrule
\end{tabular}
\label{tab:case_results_1200}
\end{table}

\section{Conclusions}

This paper showed how uncertainties in orbit determination and trajectory prediction affects imaging opportunities for Earth observing satellite systems. Position errors of \SI{1}{\kilo\meter} or larger were shown to produce significant differences in predicted imaging opportunity times. We introduced a Markov decision process formulation of the satellite imaging problem. It was shown to be a viable solution approach to the satellite scheduling problem, capable of online-planning with an execution time lower than that of standard approaches. However, due to the size of the state space, the MDP model could only be solved using an approximate forward search algorithm that led to sub-optimal tasking policies. While the MDP model underperformed compared to the Graph and MILP formulations when there was no trajectory prediction error, it outperformed the MILP approach once uncertainty was introduced. It also reduced the variance in the realized reward in the presence of uncertainties. The ease of implementation, robustness, and extensibility of the solution approach indicate that the MDP formulation is well suited for satellite task planning when the system is affected by uncertainties.

%

\bibliographystyle{named}
\bibliography{eddySatelliteTasking2019}

\end{document}